\documentclass[12pt]{emulateapj} 
  
\usepackage{hyperref}
\usepackage{verbatim}
\usepackage{graphicx}
\usepackage{color}
\usepackage{amssymb}
\usepackage{hhline}

\shorttitle{Keck/MOSFIRE Spectroscopy of $\lowercase{z}=$ 7--8 Galaxies}
\shortauthors{Song et al.}

\newcommand{\Msol}{$M_{\odot}$}

\newcommand{\OIII}{[O\,{\sc iii}]}
\newcommand{\OII}{[O\,{\sc ii}]}

\newcommand{\CIII}{C\,{\sc iii}]}
\newcommand{\CII}{[C\,{\sc ii}]}

\begin{document}
\slugcomment{Submitted to the ApJ}

\title{Keck/MOSFIRE Spectroscopy of $\lowercase{z}=$ 7--8 Galaxies: L\lowercase{y}$\alpha$ Emission from a Galaxy at $\lowercase{z} = 7.66$} 
\author{Mimi Song\altaffilmark{1},
	Steven L. Finkelstein\altaffilmark{1},
	Rachael C. Livermore\altaffilmark{1},
	Peter L. Capak\altaffilmark{2},
	Mark Dickinson\altaffilmark{3},
	and 
	Adriano Fontana\altaffilmark{4}
	}
\email{mmsong@astro.as.utexas.edu}

\altaffiltext{1}{Department of Astronomy, The University of Texas at Austin, Austin, TX 78712, USA}
\altaffiltext{2}{Spitzer Science Center, 314-6 Caltech, Pasadena, CA 91125, USA}
\altaffiltext{3}{National Optical Astronomy Observatory, Tucson, AZ 85719, USA}
\altaffiltext{4}{INAF -- Osservatorio Astronomico di Roma, via  Frascati 33, I-00040 Monteporzio, Italy}

\begin{abstract}

We report the results from some of the deepest Keck/Multi-Object Spectrometer For Infra-Red Exploration data yet obtained for candidate $z \gtrsim 7$ galaxies. Our data show one significant line detection with 6.5$\sigma$ significance in our combined 10 hr of integration which is independently detected on more than one night, thus ruling out the possibility that the detection is spurious. The asymmetric line profile and non-detection in the optical bands strongly imply that the detected line is Ly$\alpha$ emission from a galaxy at $z$(Ly$\alpha)=7.6637 \pm 0.0011$, making it the fourth spectroscopically confirmed galaxy via Ly$\alpha$ at $z>7.5$. This galaxy is bright in the rest-frame ultraviolet (UV; $M_{\rm UV} \sim -21.2$) with a moderately blue UV slope ($\beta=-2.2^{+0.3}_{-0.2}$), and exhibits a rest-frame Ly$\alpha$ equivalent width of  EW(Ly$\alpha$) $\sim 15.6^{+5.9}_{-3.6}$ \AA. The non-detection of the 11 other $z \sim$ 7--8 galaxies in our long 10 hr integration, reaching a median 5$\sigma$ sensitivity of 28 \AA\ in the rest-frame EW(Ly$\alpha$), implies a 1.3$\sigma$ deviation from the null hypothesis of a non-evolving distribution in the rest-frame EW(Ly$\alpha$) between $3<z<6$ and $z=$ 7--8. Our results are consistent with previous studies finding a decline in Ly$\alpha$ emission at $z>6.5$, which may signal the evolving neutral fraction in the intergalactic medium at the end of the reionization epoch, although our weak evidence suggests the need for a larger statistical sample to allow for a more robust conclusion.

\end{abstract}

\section{Introduction}

The Ly$\alpha$ emission line is a unique tool as the line properties encode information about the scattering medium through which the photons have passed.
During the past few years, in the present absence of a sensitive 21 cm signal from reionization, investigating the redshift evolution of the ``Ly$\alpha$ fraction'', the fraction of Lyman-break galaxies (LBGs) which exhibit strong Ly$\alpha$ emission, has served as a valuable and feasible means of providing constraints on the ionization state of the intergalactic medium (IGM). 
Spectrosopic follow-up of LBGs has revealed that the Ly$\alpha$ fraction (typically defined as LBGs with rest-frame Ly$\alpha$ EW $> 25$ \AA) steadily increases from $z=3$ to $z=6$, reaching $\sim$50\% for faint galaxies ($M_{\rm UV} > -20.25$) at $z \sim 6$ \citep{stark10,stark11}.
At higher redshifts of $z \sim 7$, however, initial expectations and attempts based on an extrapolation of the trend of the increasing Ly$\alpha$ fraction seen at lower redshifts 
found a reverse of the trend, showing only 20\%--30\% of faint galaxies with Ly$\alpha$ emission \citep[e.g.,][]{fontana10, pentericci11, ono12, schenker12, finkelstein13, pentericci14}.
This steep decrease beyond $z \sim 6$ is in line with measurements of Gunn--Peterson troughs \citep{gunn65} in the spectra of distant quasars \citep{fan06}, which signal the (near) completion of reionization by $z \sim 6$.

Several attempts have been made to interpret the observed drop in the Ly$\alpha$ fraction in connection with the neutral fraction of the IGM or different models of reionization.
Earlier works suggested, assuming that the observed drop in the Ly$\alpha$ fraction from $z \sim 6$ to $z \sim 7$ is entirely driven by the change in the IGM transmission, that it requires a steep increase in the volume-averaged neutral fraction of $\Delta x_{\rm H\,{\sc I}}>$ 0.4--0.5 over $\Delta z=1$ \citep{dijkstra11, pentericci11}.
Alternatives have subsequently been proposed that account for the possibility of other sources of Ly$\alpha$ attenuation which alleviate the amount of the required increase in the neutral fraction. 
For example, \citet{dijkstra14} suggested that the change in the intrinsic physical properties of galaxies such as an increase in the escape fraction of ionizing photons can explain the observed drop with a mild increase in the neutral fraction of $\Delta x_{\rm H\,{\sc I}} =$ 0.1--0.2, and \citet{bolton13} argued that the rise of the neutral fraction of only $\Delta x_{\rm H\,{\sc I}} =$ 0.1 by $z=7$ is sufficient when  accounting for self-shielding absorption systems (Lyman limit systems; LLSs) in the IGM, which are expected to be abundant near the end of reionization (though see \citealt{mesinger15}).
At $z \sim 7$, a sufficient sample has been assembled to start discerning between `patchy' and `smooth' models of Ly$\alpha$ attenuation. 
\citet{pentericci14} found from a compilation of observations at $z \sim 7$ that the `patchy' model of Ly$\alpha$ attenuation (which does not necessarily literally mean a patchy reionization process but may instead signal the abundant LLSs; \citealt{mesinger15}), is favored over the `smooth' attenuation model. 
Although the interpretation is not straightforward, these studies all highlight 
the potential of studying the Ly$\alpha$ fraction as a valuable probe of reionization.


\begin{deluxetable*}{lccccccccc}
\tabletypesize{\scriptsize}
\tablecaption{\label{tab:target} Summary of $z_{\rm phot}=$ 7--8 candidates observed with MOSFIRE}
\tablewidth{0pt}
\tablehead{\colhead{ID\,\tablenotemark{a}} & \colhead{R.A.} & \colhead{decl.} & \colhead{$J_{125}$} & \colhead{$H_{160}$} & \colhead{$M_{\rm UV}$} & \colhead{$z_{\rm phot}$} & \colhead{$z_{\rm phot}$ 68\% C.L.\tablenotemark{b}} & \colhead{$p(z)_{Y{\rm\,band}}$\tablenotemark{c}} & \colhead{EW$_{\rm Ly\alpha}$\tablenotemark{d}} \\
\colhead{$ $} & \colhead{(J2000)} & \colhead{(J2000)} & \colhead{} & \colhead{} & \colhead{} & \colhead{} & \colhead{} & \colhead{} & \colhead{(\AA)} 
}
\startdata
z8\_GSD\_17938    & 3:32:49.94   & $-$27:48:18.1 &  25.7 &    25.7 &   $-$21.6 &    8.07  &  $[7.87, 8.37]$  &    0.70 & $<$ 12 \\
z7\_GSD\_10175    & 3:32:50.48   & $-$27:46:56.0 & 25.7 &    25.6 &   $-$21.2 &    6.93  &  $[6.14, 7.22]$  &    0.37 & $<$ 15 \\
z7\_GSD\_12816    & 3:32:44.89   & $-$27:47:21.8 & 26.9 &    27.2 &   $-$20.2 &    6.81  &  $[6.02, 7.20]$  &    0.32 & $<$ 45 \\
z7\_MAIN\_2852    & 3:32:42.56   & $-$27:46:56.6 & 26.0 &    26.0 &   $-$20.9 &    6.85  &  $[6.75, 6.93]$  &    0.08 & $<$ 25 \\
z7\_MAIN\_4005    & 3:32:39.55   & $-$27:47:17.5 &  26.5 &    26.5 &   $-$20.7 &    7.55  &  $[6.30, 7.55]$  &    0.53 & $<$ 27 \\
z7\_MAIN\_3474    & 3:32:38.80   & $-$27:47:07.2 &  27.0 &    27.0 &   $-$20.0 &    7.41  &  $[7.08, 7.54]$  &    0.92 & $<$ 55 \\
z8\_GSD\_2135     & 3:32:42.88   & $-$27:45:04.3 &  26.9 &    26.8 &   $-$20.2 &    7.76  &  $[1.84, 8.05]$  &    0.49 & $<$ 39 \\
z7\_GSD\_568      & 3:32:40.69   & $-$27:44:16.7 & 26.9 &    26.8 &   $-$20.1 &    7.20  &  $[6.62, 7.45]$  &    0.62 & $<$ 35 \\
z7\_GSD\_431      & 3:32:40.26   & $-$27:44:09.9 & 26.6 &    26.7 &   $-$20.4 &    7.37  &  $[6.66, 7.71]$  &    0.70 & $<$ 28 \\
z7\_GSD\_1273     & 3:32:36.00   & $-$27:44:41.7 & 26.5 &    26.5 &   $-$20.4 &    6.86  &  $[6.66, 7.05]$  &    0.30 & $<$ 31 \\
z7\_GSD\_3811     & 3:32:32.03   & $-$27:45:37.1 & 25.8 &    25.9 &   $-$21.2 &    7.42  &  $[6.71, 7.62]$  &    0.73 & \,\,\,$<$ 15\,\tablenotemark{e} \\
z7\_ERS\_12098    & 3:32:35.44   & $-$27:42:55.1 & 26.3 &    26.3 &   $-$20.7 &    7.17  &  $[6.23, 7.25]$ &    0.49 & $<$ 23 
\enddata
\tablenotetext{a}{IDs from \citet{finkelstein15b}.}
\tablenotetext{b}{68\% confidence level in photometric redshift.}
\tablenotetext{c}{Integral of $p(z)$ over the MOSFIRE $Y$-band spectral coverage.}
\tablenotetext{d}{Median 5$\sigma$ rest-frame EW limit of Ly$\alpha$ calculated using the 5$\sigma$ limiting line flux for each object (see Section \ref{sec:simul}), regardless of line detection.}
\tablenotetext{e}{Note that the EW of the Ly$\alpha$ emission detected in z7\_GSD\_3811 is $15.6^{+5.9}_{-3.6}$ \AA, as reported in Table \ref{tab:3811}.}
\end{deluxetable*}


Because Ly$\alpha$ is redshifted into the near-infrared, 
pushing the study of 
Ly$\alpha$ emission to a higher redshift of $z > 7$ had been relatively slow. 
However, the advent of a new generation of ground-based near-infrared spectrographs with 
multiplexing capability and increased sensitivity has been changing the game by enabling 
more systematic searches for Ly$\alpha$ emission in $z \gtrsim 7$ galaxies. 
However, the current sample at $z>7$ lacks the statistical power to discern between the two models of Ly$\alpha$ attenuation \citep[e.g.,][]{tilvi14}, as the required sample size is predicted to be at least several tens \citep{treu12}.

As expected, previous attempts in the search for Ly$\alpha$ emission at $z>7$ 
have revealed that spectroscopically confirming galaxies at $z>7$ via Ly$\alpha$ is challenging,  
yielding, in addition to two galaxies confirmed via Ly$\alpha$-break and/or dust continuum \citep{watson15,oesch16}, only 10 spectroscopically confirmed galaxies via Ly$\alpha$ so far (\citealt{vanzella11, ono12, schenker12, schenker14, shibuya12, finkelstein13, oesch15, roberts-borsani15, zitrin15}; see review in \citealt{finkelstein15a}), and only four at $z>7.5$, possibly due to an increased neutral fraction in the IGM. 
Despite these challenges, spectroscopic follow-up of galaxy candidates at these high redshifts, either yielding detections or non-detections, 
is valuable toward building up a statistical sample that is large enough to constrain the reionization process as well as studying in detail the physical properties of galaxies via further follow-up observations, and is thus being actively pursued. 

This paper extends such previous and on-going attempts. 
In this study, we report Ly$\alpha$ emission from a galaxy at $z=7.66$ in the Great Observatories Origins Deep Survey South \citep[GOODS-S;][]{giavalisco04} field. This is from a very deep spectroscopic follow-up campaign of $z \sim$ 7--8 galaxy candidates with the 
Multi-Object Spectrometer For Infra-Red Exploration \citep[MOSFIRE;][]{mclean12} on the Keck I 10 m telescope, where we push the 
median 5$\sigma$ limiting sensitivity in line flux down to $\sim 5 \times 10^{-18}$ erg s$^{-1}$ cm$^{-2}$ between sky lines.
Although limited by the small number of observed galaxies, we discuss the implications of our results in the context of the evolution of the Ly$\alpha$ visibility.

This paper is organized as follows. Section \ref{sec:data} describes our target selection, deep spectroscopic observations with MOSFIRE, and data reduction. Section \ref{sec:results} and \ref{sec:sedfit} present the results from our spectroscopy and our stellar population modeling, respectively. The implication of our observations on the Ly$\alpha$ visibility is presented in Section \ref{sec:simul}. The discussion and summary follow in Section \ref{sec:discussion}. 
Throughout the paper, we adopt a concordance $\Lambda$CDM cosmology with $H_0 = 70$ km s$^{-1}$ Mpc$^{-1}$, $\Omega_M = 0.3$, and $\Omega_{\Lambda} = 0.7$.  
We use the AB magnitude system \citep{oke83} and a \citet{salpeter55} initial mass function (IMF) between 0.1 \Msol\, and 100 \Msol. We refer to the {\it Hubble Space Telescope} ({\it HST}) bands F435W, F606W, F775W, F814W, F850LP, F098M, F105W, F125W, F140W, and  F160W as \textit{B$_{435}$, V$_{606}$, i$_{775}$, I$_{814}$, z$_{850}$, Y$_{098}$, Y$_{105}$, J$_{125}$, JH$_{140}$, {\rm and} H$_{160}$}, respectively. All quoted uncertainties are at 68\% confidence intervals.

\section{Data}\label{sec:data}

\subsection{\textit{HST} Data and Sample Selection}

The targets were selected in the GOODS-S field from the parent sample from \citet{finkelstein15b}. 
The parent sample was selected via photometric redshifts, which were estimated with {\tt EAZY} \citep{brammer08}, 
using the {\it HST} data set from the Cosmic Assembly Near-infrared Deep Extragalactic Legacy Survey \citep[CANDELS;][]{grogin11, koekemoer11} which incorporates all earlier imaging data over the field as described by \citet{koekemoer11,koekemoer13}.
The MOSFIRE slit design was prepared using the MAGMA configurable slit unit (CSU) design tool.  This tool takes as an input a list of objects, along with relative priorities.  Our priority scheme was based on two quantities: the $J_{125}$-band magnitude of the source and the fraction of the source's redshift probability distribution function (PDF; $p(z)$) of which Ly$\alpha$ would be encompassed by the MOSFIRE $Y$ band (7.0 $\lesssim z \lesssim$ 8.2).  We first assigned an initial priority based on the continuum magnitude, and then prioritized galaxies within that continuum magnitude bin by the normalized redshift integral.  In this way, for two galaxies with similar redshift PDFs, the higher priority would go to the brighter one, while a faint galaxy with $z_{\rm phot} \sim$ 7.5 would be prioritized over a bright galaxy with $z_{\rm phot} \sim$ 6.0.
In sum, we targeted 12 (8) galaxy candidates with $z_{\rm phot}=$ 6.8--8.2 (7.0--8.2). 
Of these, six galaxies have more than half of their redshift PDF placing Ly$\alpha$ in the MOSFIRE $Y$ band.
The rest of slits in the mask were assigned to 18 galaxy candidates at lower redshifts of $z_{\rm phot}=$ 4--6 and one relatively bright star to monitor transparency and pointing accuracy.
The median rest-frame absolute UV magnitude ($M_{\rm UV}$) of our targets (assuming they are at their photometric redshifts) is $-20.4$ for the $z=$ 7--8 sample, ranging from $-21.6$ to $-20.0$. The median $H_{160}$-band magnitude is 26.5, ranging [25.6--27.2]. The full list of our $z=$ 7--8 sample is tabulated in Table \ref{tab:target}.

\subsection{MOSFIRE Y-band Observation}\label{sec:obs}

Observations were taken with MOSFIRE on the Keck I telescope over 4 nights during January 11 and January 13--15, 2015. We used the $Y$-band filter, to search for Ly$\alpha$ emission at $7.0 < z < 8.2$, with a 0\farcs7 slit width correpsonding to a spectral resolution of $\sim$3 \AA\ ($R=3500$).
Most of the data were taken with 180 s exposures per frame, except that for the data taken on one night (January 15; for a total of 0.9 hr integration time) 60 s exposures per frame were used.
We adopted an ABBA dither pattern with an $\pm$ 1\farcs25 offset along the slit for sky subtraction.
The seeing measured from the star placed on a slit was in the range of 0\farcs6--0\farcs9, with a median/mean of 0\farcs7.
In total, we obtained a total on-source integration time of $\sim$10 hr (from 2.8 hr (January 11) + 3.2 hr (January 13) + 3.2 hr (January 14) + 0.9 hr (January 15)), among which $\sim$7.3 hr was obtained in good conditions. 
These observations are among the deepest observations ever taken for $z \gtrsim 7$ galaxies.

\subsection{Data Reduction}

Data reduction was performed with the public MOSFIRE data reduction pipeline (DRP; version 2015A), in which flat fielding, wavelength calibration, sky subtraction, and rectification were performed to create two-dimensional (2D) spectra with a spectral resolution of 1.09 \AA\ pixel$^{-1}$ and a spatial resolution of 0\farcs18 pixel$^{-1}$. 
Upon monitoring the centroid of the slit star in each raw frame, we identified a $\sim$1 pixel hr$^{-1}$ drift along the slit, which was also noted by several other studies \citep[e.g.,][]{kriek15, oesch15}. We thus split the data on each night into $\sim$1 hr chunks and reduced them seperately, to prevent loss of signal due to this drift. 

Following this, analysis was done using our custom software.
From the 2D spectrum created by the pipeline, we combined the data from the four nights by generating final inverse-variance-weighted stacks for each object, following \citet{gawiser06}.
Spatial offsets between data chunks due to the drift were accounted for when combining data based on the centroids of the slit star.
We extracted one-dimensional (1D) spectra at the expected position of each source with a width of 1\farcs3 (about a 1.8$\times$ the median Gaussian FWHM), 
using an optimal extraction algorithm described in \citet{horne86}. This extraction scheme is similar to inverse-variance weighting, but additional weight is given for each spatial pixel based on the expected spatial profile for each source (which is a Gaussian for our unresolved sources), reducing statistical noise in the extracted spectra compared to a simple boxcar extraction scheme.


\begin{deluxetable}{ll}
\tabletypesize{\scriptsize}
\tablecaption{\label{tab:3811} Summary of z7\_GSD\_3811}
\tablewidth{0pt}
\tablehead{
\multicolumn{2}{c}{Emission Line Properties}
}
\startdata

$F_{\rm Ly\alpha}$ (10$^{-18}$ erg s$^{-1}$ cm$^{-2}$) & $5.5 \pm 0.9$ ($\pm 1.7$) \\
Signal-to-noise Ratio & 6.5  \\
EW$_{\rm Ly\alpha}$ (\AA)\,\tablenotemark{a}  & 15.6$^{+5.9}_{-3.6}$ ($\pm 4.7$) \\
$z_{\rm Ly\alpha}$ & $7.6637 \pm 0.0011$ \\
$\sigma_{\rm blue}$ (\AA)\,\tablenotemark{b} & $0.33^{+5.51}_{-0.32}$ \\
$\sigma_{\rm red}$ (\AA)\,\tablenotemark{b} & $6.49^{+0.32}_{-4.76}$ \\
FWHM$_{\rm red}$ (\AA)\,\tablenotemark{c} & $15.0 \pm 2.7$ \\ 
$\sigma_{\rm red}$ (km s$^{-1}$)\,\tablenotemark{d}  & $180 \pm 30$ \\

\cutinhead{Physical Properties} \\

log $M_*$ (\Msol) & $9.3^{+0.5}_{-0.4}$ \\
UV slope $\beta$\,\tablenotemark{e} & $-2.2^{+0.3}_{-0.2}$  \\
$M_{\rm UV}$ &  $-21.22^{+0.06}_{-0.10}$ \\ 
$E(B-V)$ & $0.06^{+0.10}_{-0.04}$ \\
SFR$_{\rm UV,obs}$ (\Msol\,yr$^{-1}$) &   $19^{+2}_{-1}$ \\
SFR$_{\rm UV,corr}$ (\Msol\,yr$^{-1}$)\,\tablenotemark{f} &   $33^{+56}_{-9}$ 

\enddata
\tablecomments{Listed in parentheses are systematic uncertainties.}
\tablenotetext{a}{Rest-frame equivalent width of Ly$\alpha$.}
\tablenotetext{b}{Observed line width of the blue and red side of the asymmetric Gaussian line profile, respectively.}
\tablenotetext{c}{Observed FWHM of the red side of the line.}
\tablenotetext{d}{Line-of-sight velocity dispersion inferred from the red side of the line. Corrected for instrumental resolution.}
\tablenotetext{e}{UV slope obtained in a same way as to \citet{finkelstein12}, by fitting the wavelength window in the 1300--2600 \AA\ region defined by \citet{calzetti94} of the best-fit SPS model as a power law.}
\tablenotetext{f}{Dust-corrected SFR from the observed rest-frame UV magnitude and $E(B-V)$ obtained from the SPS model, assuming the \citet{calzetti00} extinction law and the \citet{kennicutt98} conversion.}
\end{deluxetable}


\begin{figure*}[t]
  \epsscale{1.0}
\plotone{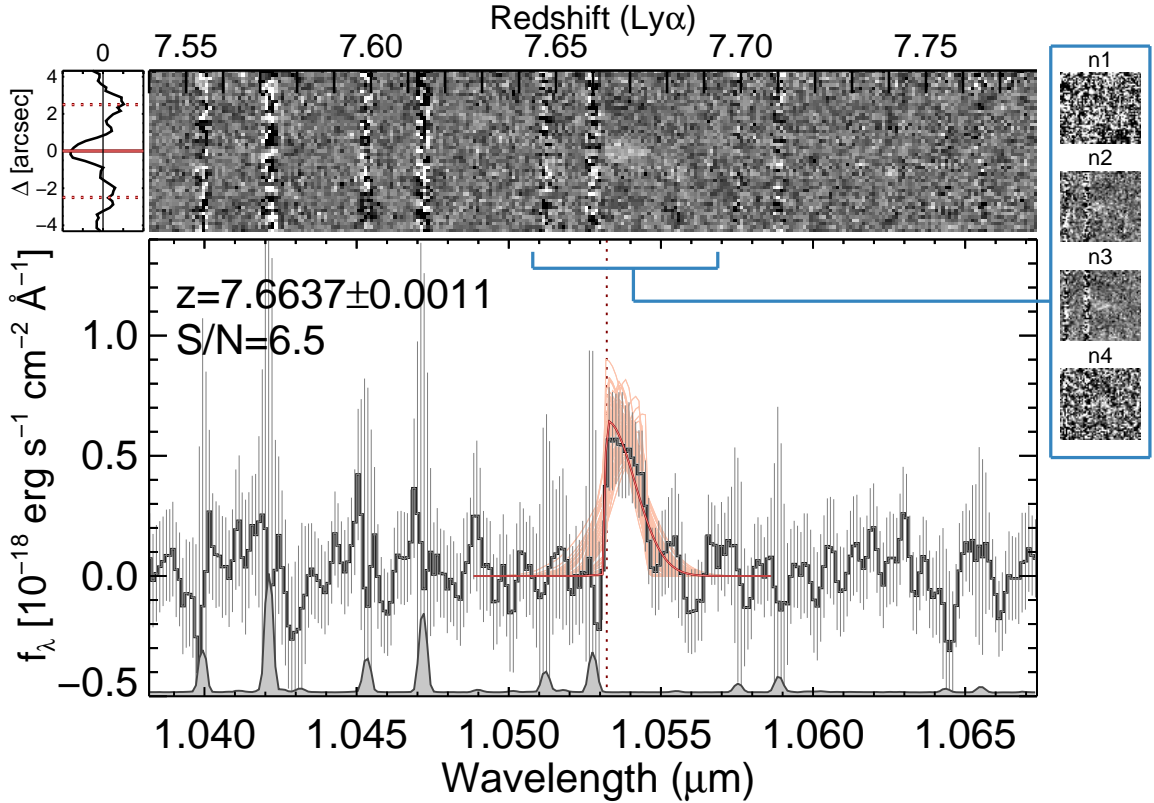}
  \epsscale{1.0}
  \caption{\label{fig:spec}
 MOSFIRE $Y$-band 2D ($top$) and 1D ($bottom$) stacked spectra for the object with detected emission (z7\_GSD\_3811), showing a clear asymmetric line profile characteristic of Ly$\alpha$ emission. 
The displayed 1D spectrum was smoothed by the instrumental resolution ($\sim$3 \AA). 
The best-fit asymmetric Gaussian curve and the line centroid are overplotted as the red thick solid curve and red dotted vertical line, respectively.
The red thin curves are 100 Monte Carlo fits. 
The gray-shaded region near the bottom of the 1D spectrum shows a scaled sky spectrum.
 Also shown on the upper left corner is the 1D spectrum of the emission extracted along the spatial direction with inverse-variance weighting over the extraction width of the FWHM of the line. The red solid line and two red dotted lines overplotted are the expected spatial location of the postive peak and two negative peaks, respectively.
We show in the blue box on the right side that the emission line is independently detected on all nights (n2, n3, n4) except in n1 which suffered from poor conditions, 
indicating that the chance of the detection being a spurious one is negligible.
 }
 \end{figure*}
 

Correcting for telluric absorption was done using the \citet{kurucz93} model spectrum of the spectral type of the slit star (G5I). Absolute flux calibration was performed by comparing and scaling the spectrum to the WFC3 $Y_{105}$-band magnitude of the slit star. This procedures accounts for the slit loss, assuming our targets are point sources unresolved under the seeing FWHM of our observations, which is a good approximation given the small size of high-redshift galaxies. 

To check our flux calibration, we compared our calibration array with the total MOSFIRE $Y$-band throughput curve.\footnote{http://www2.keck.hawaii.edu/inst/mosfire/throughput.html}
We also utilized two bright continuum sources 
which were serendipitously included in our mask, to further verify our absolute flux calibration.
Taking a similar approach to that of \citet{kriek15}, we first convolved {\it HST}/{\it Y$_{105}$} images of the two sources and the slit star with a Gaussian kernel with width ${\rm FWHM}_{\rm kernel}^2={\rm FWHM}_{\rm seeing}^2-{\rm FWHM}_{\rm {\it H}_{160}}^2$, to generate the $Y$-band image under the seeing of our spectroscopic observations. Then, we calculated the fraction of light of the two sources that are within our MOSFIRE slit layout. Comparing them to the fraction of light of the star within the slit (on which our absolute flux calibration is based), we calculated the expected flux ratio between our spectroscopic data and the broadband flux (i.e., {\it HST}/{\it Y$_{105}$}) for the two sources due to the difference in the slit loss. This comparison shows that our absolute calibration (which affects our measurements of line flux and equivalent width, but not the significance of the detection) is accurate within 20\%--25\%. We thus conservatively add a 30\% systematic uncertainty in calibration in our error budget. The systematic uncertainties are indicated in Table \ref{tab:3811}, while the quoted uncertainties in the rest of the paper refer to random uncertainties. 

Finally, to make sure that the error spectrum initially obtained from the pipeline does not underestimate the noise level, we scaled the error spectrum such that the standard deviation of the signal-to-noise ratio (S/N) in the sky dominated region is unity. 
The typical scale factor was $3.0 \pm 0.1$.

\section{Results}\label{sec:results}

\subsection{Line Detection}

We visually searched for emission lines in the extracted 1D spectra as well as 2D spectra at the expected positions of our targets.
We take a conservative appoach of presenting objects for which an emission line is independently detected on more than one night, minimizing the possibility of a spurious detection. 
In other words, we regarded it as a spurious detection if the emission was detected on only one night out of four nights. This criterion yielded only one line detection among the 30 objects originally targeted, 
at $\lambda_{\rm obs}= 10532.2 \pm 1.3 $ \AA, and with $6.5 \sigma$ significance. The rest remained undetected ($< 3\sigma$).
Figure \ref{fig:spec} shows the 1D and 2D spectra of the object with emission, z7\_GSD\_3811. 
The emission is detected on more than one night at the same spatial and spectral location, with two negative peaks at the expected position from the adopted dithering pattern, ensuring that the line is real and not spurious. 

Normally, we expect an asymmetric line profile with a sharp blue edge and gradually declining red tail for Ly$\alpha$ emission at high redshift due to absorption by neutral hydrogen in the interstellar and intergalactic medium. 
However, most of the proposed Ly$\alpha$ detections in other $z \gtrsim 7$ candidates have not shown highly significant evidence for asymmetry, possibly due to the low S/N for most of the detections. 
We find that our detected line displays an asymmetric line profile, making this object one of the first notable detections of asymmetry for a $z>7$ Ly$\alpha$ line candidate.
However, the significance is not strong due to the low S/N: 
the Gaussian line width on the blue and red side of the line is $0.33^{+5.51}_{-0.32}$ \AA\ and $6.49^{+0.32}_{-4.76}$ \AA, respectively. 
Due to the vicinity of a sky line located blueward of the line, the uncertainty in the line width on the blue side of the line ($\sigma_{\rm blue}$) is large, yielding a weak constraint on
the ratio between the line width on the red and blue side
($\sigma_{\rm red}/\sigma_{\rm blue}= 19.5^{+0.2}_{-19.3}$).

\begin{figure}[t]
\epsscale{1.385}\plotone{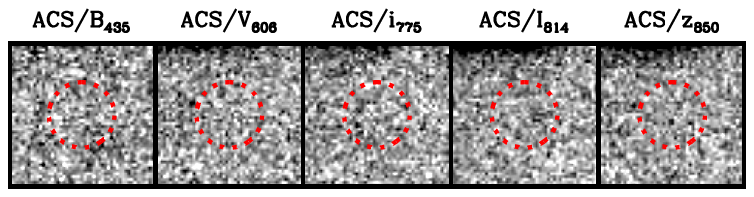}
\vspace{-23pt}
\epsscale{1.385}\plotone{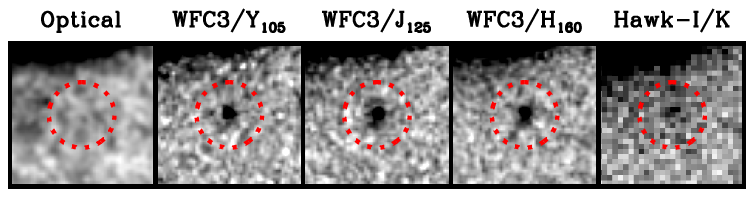}
\vspace{-33pt}
 \epsscale{1.2}\plotone{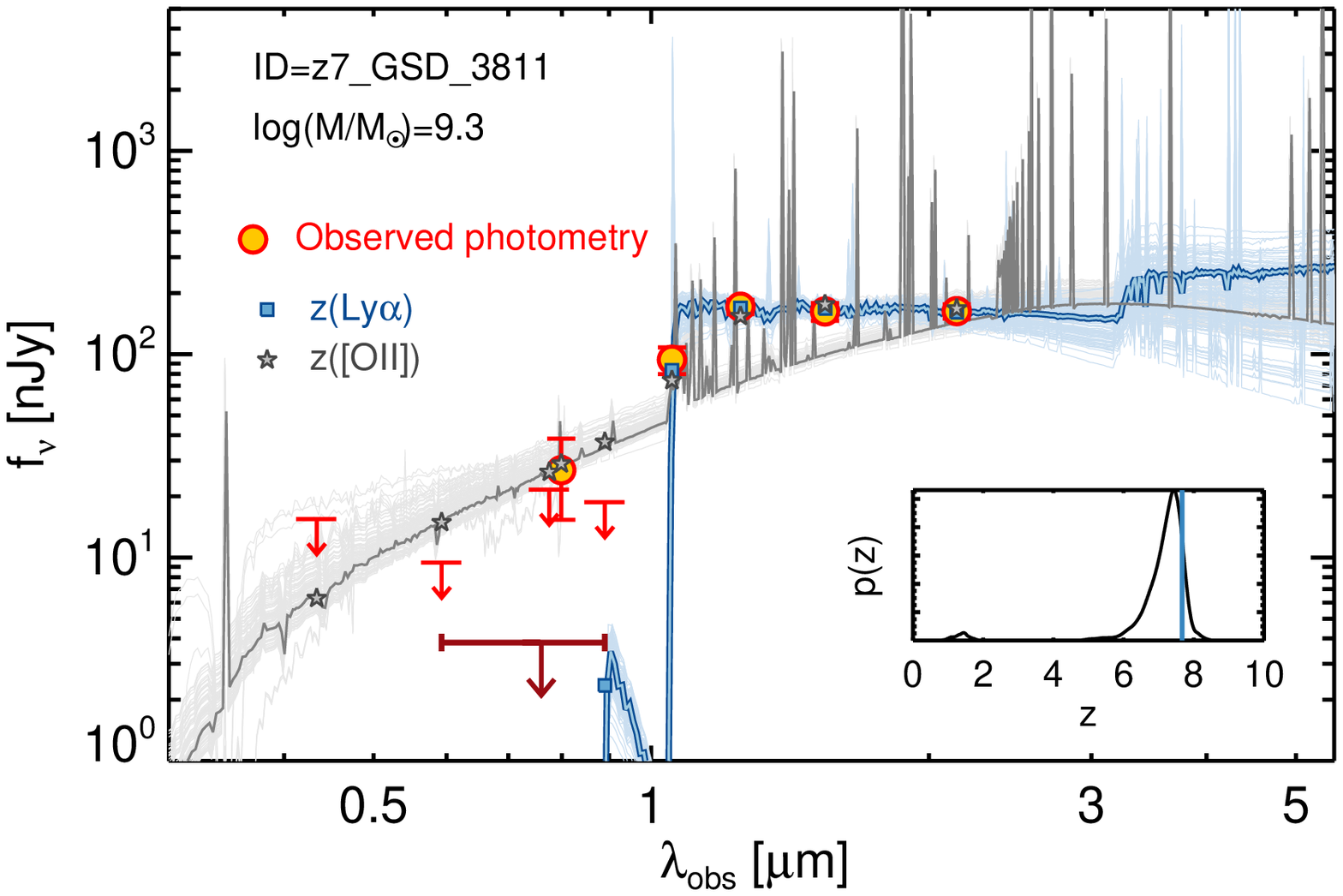}
  \epsscale{1.0}
  \caption{\label{fig:sed}
$Top$: postage stamp images of z7\_GSD\_3811 showing, from upper left to lower right, 
{\it HST}/ACS {\it B$_{435}$}, {\it V$_{606}$}, {\it i$_{775}$}, {\it I$_{814}$}, {\it z$_{850}$}, 
{\it HST}/ACS stack ({\it V$_{606}$}+{\it i$_{775}$}+{\it I$_{814}$}+{\it z$_{850}$} bands), {\it HST}/WFC3 $Y_{\rm 105}$, $J_{\rm 125}$, $H_{\rm 160}$, and VLT/Hawk-I $K$ band. 
All stamp images are 3\arcsec\ on a side, north up, east to the left.
$Bottom$: the observed SEDs (orange circles) and the best-fit SPS model and model bandpass-averaged fluxes (blue curve and blue squares) for z7\_GSD\_3811.
For non-detections, we list 1$\sigma$ upper limits (downward arrows).$^{7}$
The dark red downward arrow represents the 1$\sigma$ upper limit for
the optical stack ({\it V$_{606}$}+{\it i$_{775}$}+{\it I$_{814}$}+{\it z$_{850}$} bands). 
The best-fit SPS model and model fluxes under the alternative interpretation for the detected line (i.e., \OII\ doublet at $z=1.83$) are also shown as the gray curve and gray stars.
The thin light-colored lines are 100 Monte Carlo fits, showing that the low-$z$ solution is disfavored by the non-detection in the deep optical bands.
The inset shows the probability distribution function of photometric redshift, in good agreement with the redshift of the Ly$\alpha$ emission (blue vertical line). 
}
 \end{figure}

Assuming the line is Ly$\alpha$, the implied redshift (based on the line centroid defined as the wavelength of the peak of the Ly$\alpha$ emission) 
is $z({\rm Ly\alpha})= 7.6637 \pm 0.0011$,\footnote{Due to the IGM absorption and Ly$\alpha$ kinematics, the systemic redshift is likely to be slightly lower than the inferred redshift from the Ly$\alpha$ line. 
The systemic redshift (not corrected for IGM absorption) would be $\sim$0.01 lower than the inferred redshift for the average velocity offsets of 200--400 km s$^{-1}$ found in Ly$\alpha$ emitters and LBGs at lower redshift of $z \sim$ 2--3 \citep[e.g.,][]{song14,steidel10}.} placing it as presently the 
third most distant spectroscopically confirmed galaxy via Ly$\alpha$ and the only galaxy at $z>7$ in the GOODS-S field with a significant Ly$\alpha$ detection. 
The photometric redshift, estimated with {\tt EAZY} \citep{brammer08}, is $z_{\rm phot}=7.42^{+0.20}_{-0.71}$, 
in good agreement with the spectrosopic redshift, as shown in the inset of Figure \ref{fig:sed}.

The line-of-sight velocity dispersion, derived from the Ly$\alpha$ line width on the red side of the line and corrected for instrumental resolution, is 180 $\pm$ 30 km s$^{-1}$, 
similar to previously spectroscopically confirmed galaxies at similar redshifts \citep{oesch15,zitrin15}

We fit an asymmetric Gaussian to the line to estimate the line flux of (5.5 $\pm$ 0.9) $\times 10^{-18}$ erg s$^{-1}$ cm$^{-2}$. 
We estimated the rest-frame EW of Ly$\alpha$ emission from the observed Ly$\alpha$ flux and 
continuum flux density of the best-fit stellar population synthesis (SPS) model 
in a rest-frame 100 \AA\ box redward of the Ly$\alpha$ line (see Section \ref{sec:sedfit}).
The inferred rest-frame EW of Ly$\alpha$ emission is modest with $15.6^{+5.9}_{-3.6}$ \AA, thus this object would not be classified as an Ly$\alpha$ emitter according to the traditional criterion of  EW(Ly$\alpha$) $>$ 20 \AA. This value is also below the cutoff of EW(Ly$\alpha$) $>$ 25 \AA\ \citep{stark11} often adopted in the study of the evolution of the Ly$\alpha$ fraction 
at high redshift.


\subsection{Low-z interpretations}\label{sec:lowz}

We examined the possibility that the object is a foreground \OII$\lambda\lambda$3726, 3729, H$\beta$, \OIII$\lambda\lambda$4959, 5007, or H$\alpha$ emitter. 
First, if the detected line is H$\beta$ or one of the \OIII\ doublet, the other two lines would have been detected within our spectral coverage in regions free from sky lines. We did not find any signal at the expected wavelengths of these lines.


\begin{figure*}[t]
  \epsscale{0.57}
  \plotone{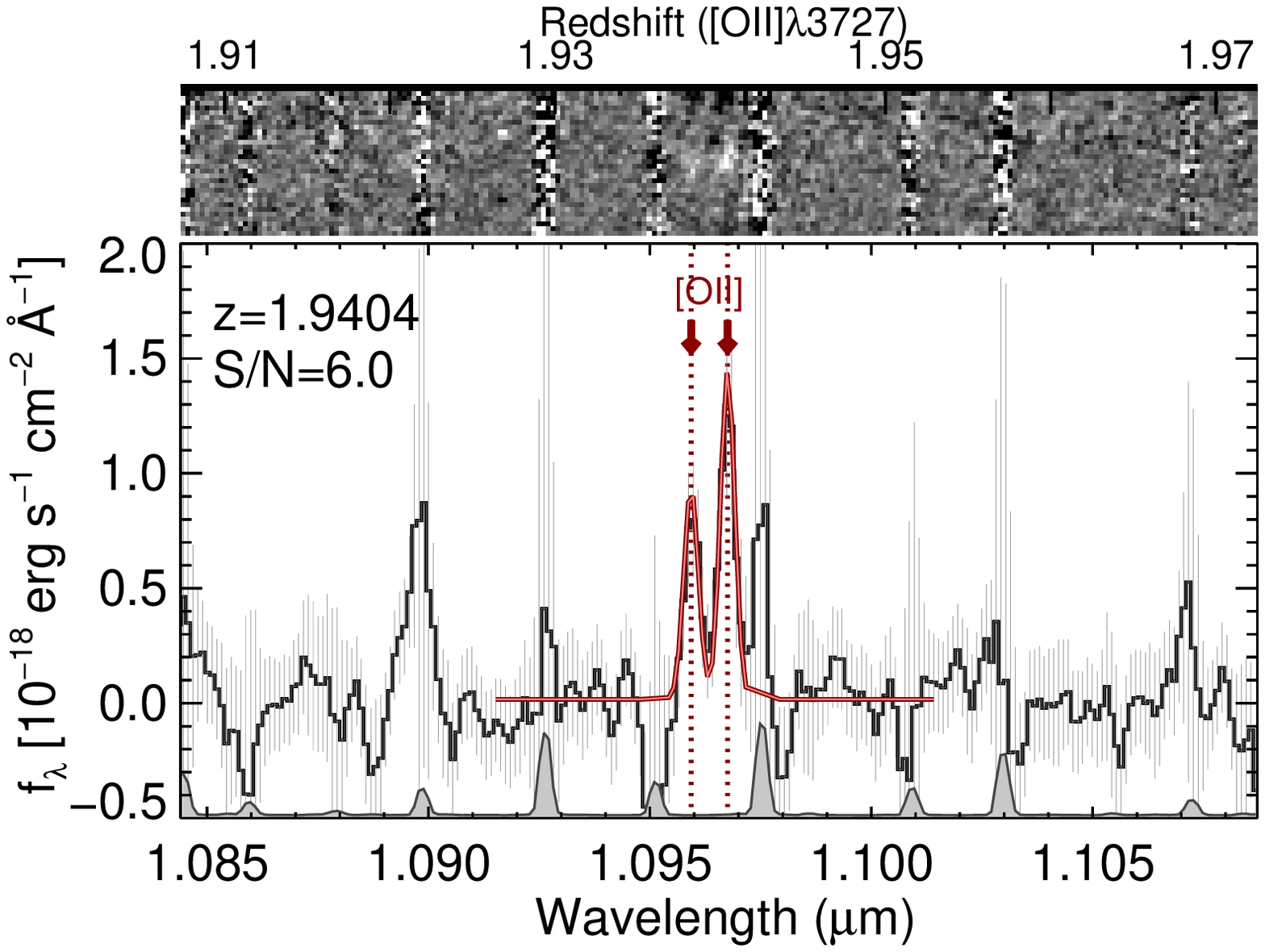}
\plotone{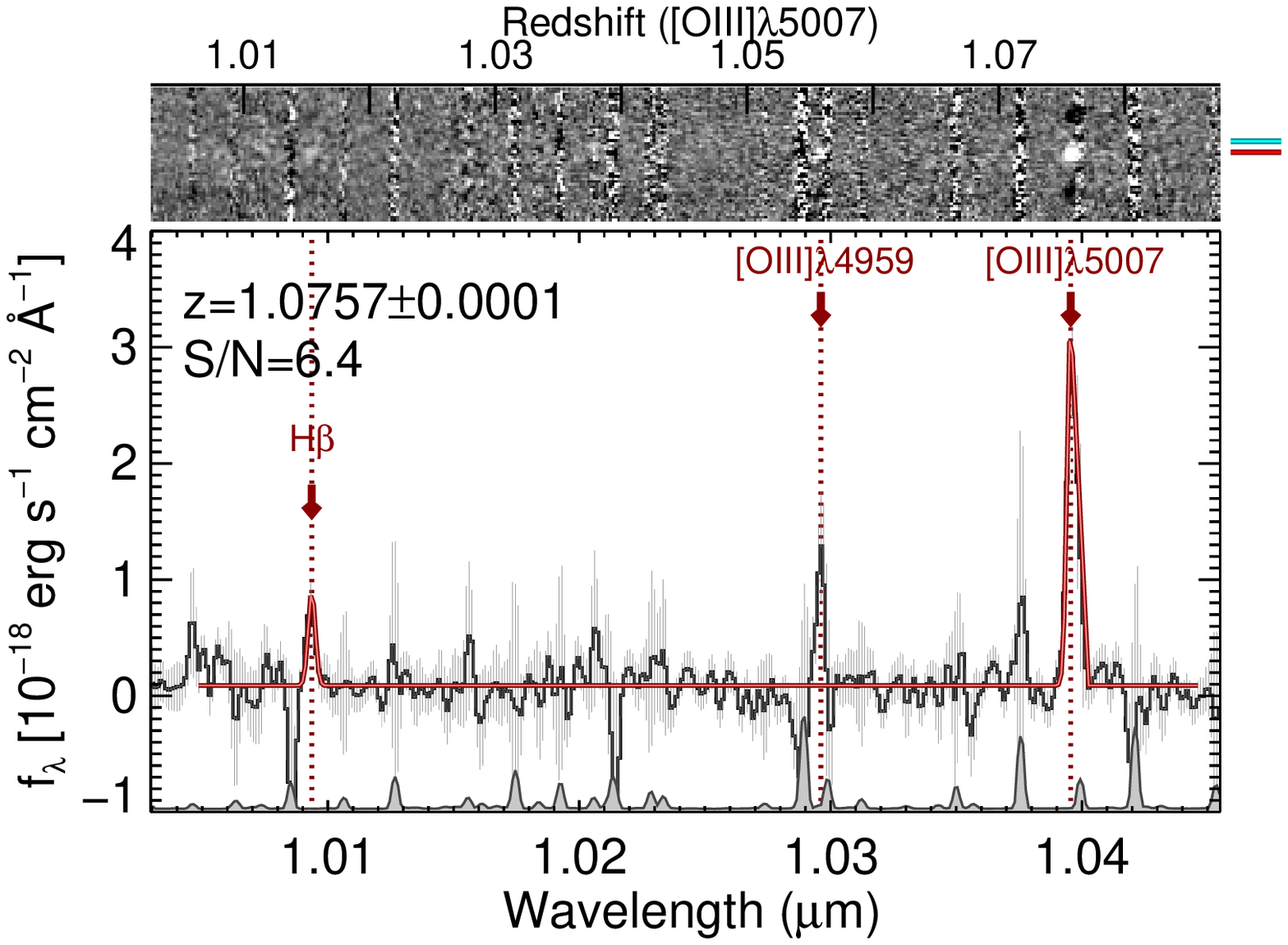}
  \epsscale{1.0}
  \caption{\label{fig:serendi}
  $Left$: an \OII\ emitter serendipitiously detected in the same mask under the same observing conditions as z7\_GSD\_3811. This source, detected at only slightly longer wavelength of $\lambda_{\rm obs}=$ 1.096 $\mu$m than z7\_GSD\_3811, exhibits a well-resolved doublet both in its 1D and 2D spectra, indicating the possibility of the line detected in z7\_GSD\_3811 being an unresolved \OII\ doublet is low. 
$Right$: detection of the H$\beta$ and \OIII$\lambda\lambda$4959,5007 doublet from a source 
close to one of our original targets. 
The cyan and red lines on the right side of the 2D spectrum mark the expected positions of our original target and the nearby source, respectively.
The position of the emission is spatially consistent with the position of the nearby source, not our original target.
}
\end{figure*}

Practically, the strong break observed between the {\it z$_{850}$} and {\it Y$_{105}$} bands (see Figure \ref{fig:sed}) rules out the possibility that the detected emission is H$\beta$, \OIII, or H$\alpha$, and leaves the only alternative possibility of the detected line being the \OII\ doublet. 
If the detected emission line is an \OII\ doublet at $z=1.83$, the spectral resolution of MOSFIRE $Y$-band grating ($\sim$3 \AA) is sufficient to resolve the doublet. 
The possibility of the detected line being one of the two peaks, however, cannot be entirely ruled out. 
If the emission is the first peak of the \OII\ doublet at $\lambda_{\rm rest}=3726$\,\AA, we would have detected the second peak (at $\lambda_{\rm rest}=3729$\,\AA) at 2--10$\sigma$ significance at wavelengths clear of sky lines. 
On the other hand, if the emission is the second peak of the doublet, the centroid of the first peak would be behind the sky line located blueward of the detected line. 
To examine these possibilities, we performed simulations in which we inserted mock lines representing either the first or second peak of the \OII\ doublet at the expected positions in the 2D spectrum. The spatial and spectral line profile of the mock line was assumed to be the same as that of the observed emission, and the flux was assigned based on the most unfavorable flux ratio that is physically allowed 
(i.e., the weakest line possible; 0.35 $<$ $f$(\OII$\lambda$3729)/$f$(\OII$\lambda$3726) $<$ 1.5;  \citealt{pradhan06}).
Our simulation results indicate that due to its low flux and broad line profile, we would not be able to completely rule out the existence of the other line of the doublet based solely on our 2D spectrum. If the detected emission is indeed one of the \OII\ doublet,  
the broad line width of the detected emission (FWHM $\sim$ 400 km s$^{-1}$) is atypical for its mass ($\log(M_*/M_{\odot})= 9.1^{+0.05}_{-0.09}$), exhibiting a factor of 3 deviation from the Tully--Fisher relation \citep{miller12}.
The line width, together with the red spectral energy distribution (SED) and lack of detection in X-rays, indicates either that \emph{if} this line is \OII\, then this galaxy likely hosts an type-2 active galactic nucleus (AGN) or that the galaxy has strong outflows. 
The direct constraint on the abundance of such population is not feasible currently at this redshift and in low-mass regime.

As discussed above, it is unlikely that the detected line is an \textit{unresolved} \OII\ doublet given the spectral resolution. 
However, since the detected line has a moderate S/N of 6.5$\sigma$, we conservatively leave this possibility open but further suggest evidence against it in Section \ref{sec:serendi} and \ref{sec:sedfit}.

\subsection{Serendipitious Line Detections at $z \sim$ 1--2}\label{sec:serendi}

In addition to the detected emission in z7\_GSD\_3811 from our targets, we identified two other emission lines in objects which serendipitiously fell in slits.

The first object (R.A. = 3:32:43.22, decl. = $-$27:47:12.9 (J2000))
shows an emission line with two peaks. 
Assuming that the detected line is an \OII\ doublet, we derived its redshift to be $z=1.94$. 
Its photometric redshift, $z_{\rm phot}=1.87^{+0.07}_{-0.08}$ \citep{dahlen13}, is in excellent agreement with the inferred \OII\ redshift, thus we conclude that the detected line is the \OII\ doublet.
This \OII\ doublet strengthens the possibility that the detected emission in z7\_GSD\_3811 is Ly$\alpha$ and not an \textit{unresolved} \OII\ doublet. The left panel of Figure \ref{fig:serendi} shows that the doublet in this object is spectrally well-resolved both in the 1D and 2D spectra, 
yet the observed wavelength and S/N are similar to those of z7\_GSD\_3811. 

The second object (R.A. = 3:32:50.48, decl. = $-$27:46:56.0 (J2000))
shows a prominent emission at $\lambda_{\rm obs}=$ 10398 \AA, which we identified as an \OIII$\lambda$5007 line (right panel of Figure \ref{fig:serendi}). The other line of the doublet (\OIII$\lambda$4959) is behind a sky line but still visible, and H$\beta$ is detected at 5.6$\sigma$. Upon close inspection, we noted that the emission has 
an offset of 4--5 pixels along the spatial axis from our original target, which corresponds to 0\farcs7--0\farcs9. 
We identified a galaxy in proximity of our original target at this distance, thus we concluded that the emission is not from our target but from a foreground galaxy at $z=1.08$.

\section{Stellar Population Modeling and Stacking Analysis}\label{sec:sedfit}

We performed a SED fitting analysis to the observed {\it HST}/Advanced Camera for Surveys (ACS; \textit{B$_{435}$, V$_{606}$, i$_{775}$, I$_{814}$, z$_{850}$}), {\it HST}/WFC3 (\textit{Y$_{105}$, J$_{125}$, H$_{160}$}), and VLT/Hawk-I $K$-band photometry of z7\_GSD\_3811, 
using the \citet{bruzual03} SPS models. Details on our modeling are described in \citet{song15}.
In addition to the {\it HST} bands originally included in the SED fitting in \citet{song15}, in this work we included the $K$-band photometry from the Hawk-I UDS and GOODS Survey \citep{fontana14} in the official CANDELS GOODS-S catalog (version 1.1).
The {\it Spitzer}/IRAC photometry was excluded from the modeling, because z7\_GSD\_3811 is unfortunately heavily contaminated by a nearby bright source in IRAC. Thus, we do not have constraints on whether this galaxy exhibits the 4.5 $\mu$m color excess due to the strong \OIII\ line falling in the 4.5 $\mu$m band 
that some other studies have reported for spectroscopically confirmed $z \sim $ 7--8 galaxies \citep{finkelstein13, oesch15, roberts-borsani15, zitrin15}. 

As discussed in Section \ref{sec:lowz}, the only alternative interpretation of the detected emission in z7\_GSD\_3811 is the \OII\ doublet. Thus, we performed the SED fitting two times with a fixed redshift, first assuming the emission is Ly$\alpha$, and then, assuming the emission is an \OII\ doublet at $z$(\OII) = 1.83.


\begin{figure*}[t]
  \epsscale{0.52}
\plotone{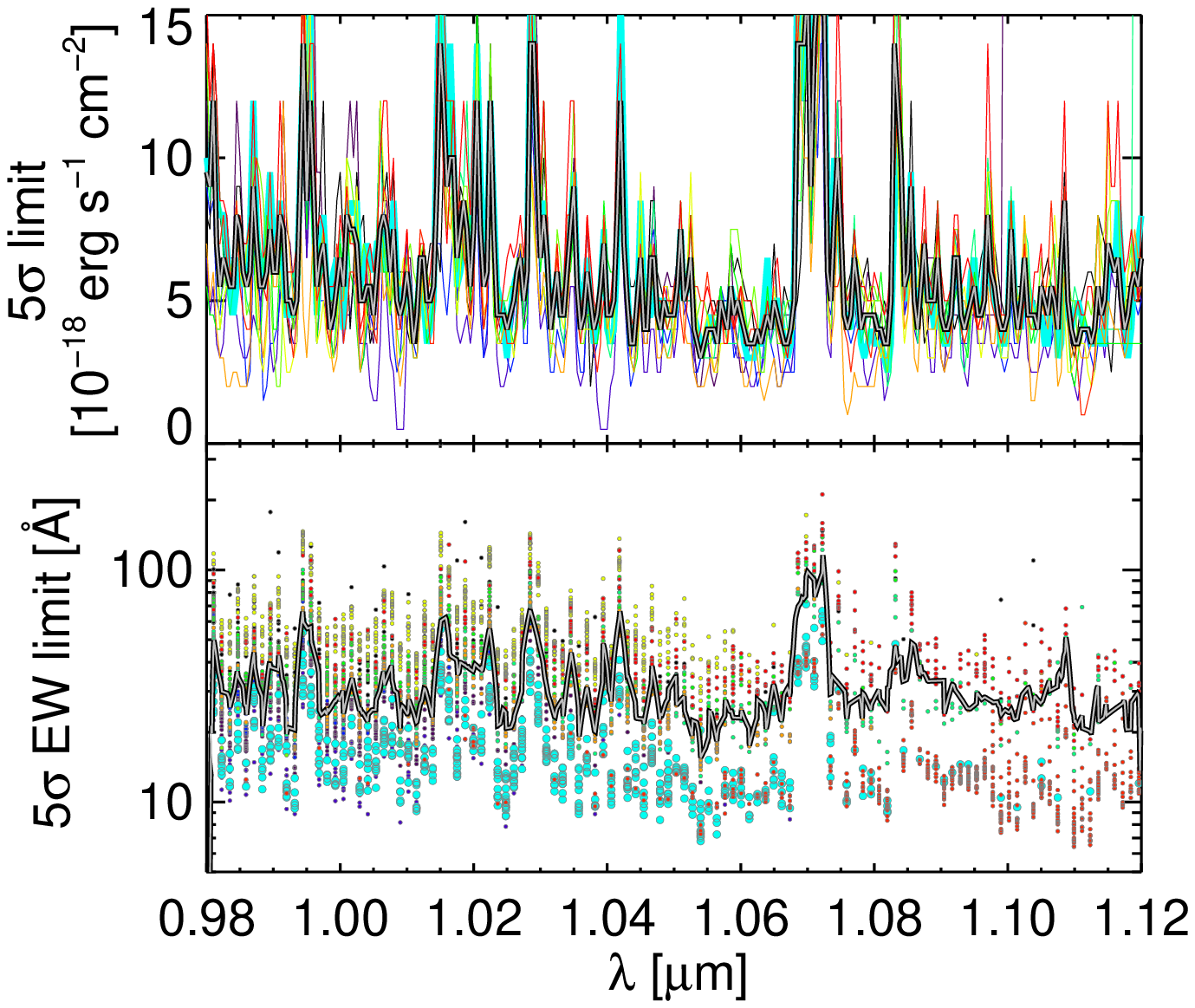}
\epsscale{0.63}
\plotone{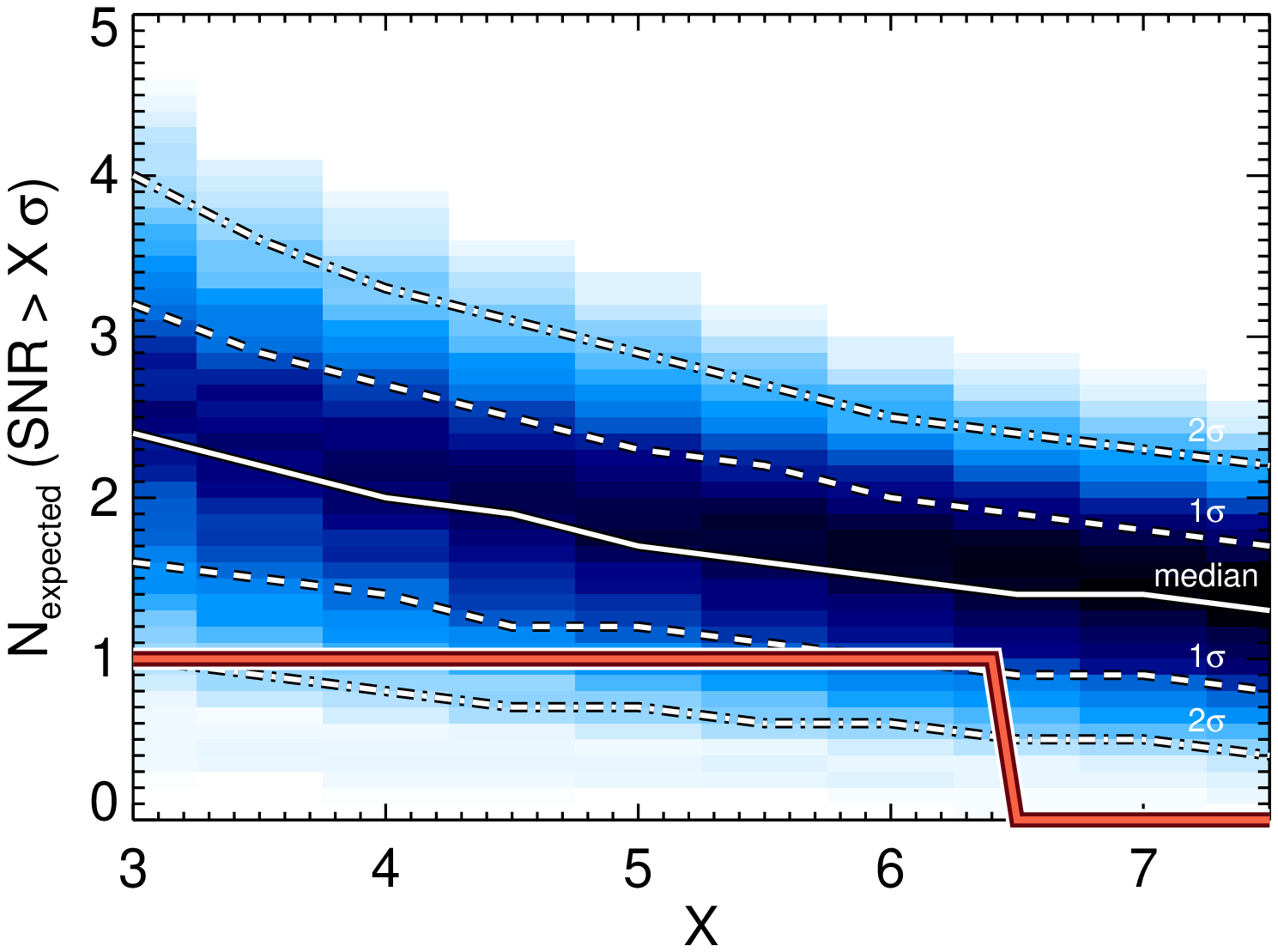}
    \epsscale{1.0}
  \caption{\label{fig:simul}
 $Upper~left$: 5$\sigma$ limiting line flux as a function of wavelength, estimated from a Monte Carlo simulation in which we inserted and recovered fake Ly$\alpha$ lines with varying line flux into our MOSFIRE spectra.
The thin lines with different colors denote the estimates for each slit, 
and the black solid line indicates the median.
$Bottom~left$: 5$\sigma$ rest-frame EW(Ly$\alpha$) limit as a function of wavelength. Each symbol denotes a trial of our Monte Carlo simulation, where different color indicates each object in our sample. The cyan circles represent the EW limits determined via Monte Carlo trials for z7\_GSD\_3811. Our $\sim$10 hr deep spectroscopy reaches a median rest-frame EW(Ly$\alpha$) of 28 \AA\ between sky lines (range = [12--55] \AA) for our $z=$ 7--8 sample.
$Right$: probability distribution of the expected number of detections for the Ly$\alpha$ line (as a function of various detection thresholds $X$) for our MOSFIRE observations, assuming no evolution with redshift in the EW(Ly$\alpha$) distribution from $3<z<6$. A darker blue color denotes higher probability. Our results of only one detection with 6.5$\sigma$ (red line) is deviated from the
null hypothesis of no EW evolution 
at 1.3$\sigma$ (for $>5\sigma$ detection, or 2$\sigma$ if we push the detection significance down to $>3\sigma$).
 }
 \end{figure*}
 

Figure \ref{fig:sed} shows the model fit and stamp images. The results of our SED fitting analysis show that the high\nobreakdashes-$z$ solution is preferred over the low-$z$ solution, albeit mildly.
For the high-z interpretation, because we did not fit bands shortwards of the Ly$\alpha$ line due to the large uncertainty in modeling the IGM attenuation, and because the source is highly contaminated by a nearby bright source in IRAC channels, only four bands (\textit{Y$_{105}$, J$_{125}$, H$_{160}$}, and $K$) were used to constrain the fit, which can be perfectly matched by SPS models with a certain combination of free parameters and nebular emission strengths, yielding $\chi^2_r \sim 0$.

For the low-$z$ interpretation, 
the non-detection in the deep optical bands\footnote{Formally, our elliptical aperture photometry yields a 2.3$\sigma$ detection in $I_{\rm 814}$ band. However, the $I_{814}$-band stamp image shows that all identifiable emissions are off-center and do not line up with near-infrared emission, indicating that they are likely background noise or from another unresolved faint source. Using a smaller, circular 0\farcs4 diameter aperture centered on the near-infrared emission, we find no detection ($< 1\sigma$) in any optical band.} and the strong break between \textit{z$_{850}$} and \textit{Y$_{105}$} of $\sim$1.8 magnitude yield the only possible solution to be a dusty low-mass ($\log(M_*/M_{\odot})= 9.1 \pm 0.1$) starburst galaxy with specific star formation rate (sSFR) of log(sSFR yr$^{-1}$)$= -7.1 \pm 0.2$. 
While the reduced chi-square of $\chi^2_r \sim 1.6$ for the low-$z$ solution indicates that it is still regarded a ``good'' fit, this low-$z$ solution is disfavored  
by non-detections in deep optical bands: 
as another measure of goodness-of-fit, we compared the distribution of normalized residuals to the standard normal distribution with ($\mu, \sigma$) = (0, 1). The comparison quantified using a Kolmogorov--Smirnov test \citep{kolmogorov33,smirnov48} indicates that the likelihood that the normalized residuals come from the normal distribution is less than  20\%, implying that the low-$z$ solution is not a preferred model for this galaxy.

To further probe the existence of any low level flux below the detection threshold of individual optical bands, we created a stack of {\it V$_{606}$}-, {\it i$_{775}$}-, {\it I$_{814}$}-, and {\it z$_{850}$}-band images. Prior to the stacking, the spatial resolution of the images were matched to that of the {\it H$_{160}$} band and the units were converted to a physical unit. Then, the stack and stack rms map were generated by inverse-variance weighting, on which the stack flux and flux error were measured within a 0\farcs4 diameter aperture using the Source Extractor package \citep{bertin96} and aperture-corrected using the ratio between the flux within a 0\farcs4 aperture and total flux measured in the {\it H$_{160}$} band. 
We quantified the background noise as the Gaussian width of the flux distribution measured from 10$^4$ randomly placed apertures of the same size used in our original photometry in source-free regions of the stacked image. We checked that the flux error (3.8 nJy) measured from Source Extractor is slightly larger than the background noise (2.6 nJy), thus conservatively took the larger one.
The stamp image and 1$\sigma$ upper limit for the flux of the stack are shown in Figure \ref{fig:sed}.
The stacking yielded no identifiable emission at the position of the source. The measured stack flux is 
$6 \sigma$ lower than the prediction from the low-$z$ solution, 
further indicating the preference for the high-$z$ interpretation of the source.

We conclude that the detected line is Ly$\alpha$. 
z7\_GSD\_3811 is a galaxy bright in the UV
with the rest-frame UV absolute magnitude of $M_{\rm UV} \sim -21.2$, about two times brighter in luminosity than the characteristic UV magnitude of the rest-frame UV luminosity function at $z=8$ of $M_{{\rm UV}, z=8}^*=-20.48$  \citep{finkelstein15a}.
Other physical properties inferred from our SED fitting analysis indicate that z7\_GSD\_3811 is a typical galaxy at $z=$ 7--8 {\it for its UV magnitude}, with a moderately blue UV slope ($\beta= -2.2^{+0.3}_{-0.2}$), dust-corrected UV-based star formation rate (SFR) of $33^{+56}_{-9}$ \Msol\, yr$^{-1}$, and stellar mass of $\log(M_*$/\Msol)$=9.3^{+0.5}_{-0.4}$. 
This galaxy was noted as 
a promising $z\gtrsim7$ candidate by several other previous {\it HST} imaging studies as well \citep{bouwens10,mclure13}.
Table \ref{tab:3811} summarizes the physical properties of z7\_GSD\_3811.

\section{L\lowercase{y}$\alpha$ Visibility}\label{sec:simul}

Even with our deep integration of 10 hr, we detected only one Ly$\alpha$ emission line with a moderate rest-frame Ly$\alpha$ EW of 16 \AA.
To put this result in context, we computed the number of detections expected from our observations, 
with the aim of placing constraints on the evolution of the Ly$\alpha$ visibility with redshift.

First, we quantified the limiting sensitivity of our observations by simulating Ly$\alpha$ lines in our MOSFIRE spectra. 
We modeled the Ly$\alpha$ line as an asymmetric Gaussian, similar to the detected line in z7\_GSD\_3811.
Then, we inserted the lines with varying line fluxes into each of the actual 1D spectra in our mask at varying positions between the MOSFIRE $Y$-band wavelength coverage (9800--11200 \AA), to find the line flux as a function of wavelength that ensures an $X$-$\sigma$ detection ($X \ge 3$).
The upper left panel of Figure \ref{fig:simul} presents the results, showing that our deep spectroscopy reaches a median 5$\sigma$ limiting sensitivity in line flux of $\sim$5 $\times 10^{-18}$ erg s$^{-1}$ cm$^{-2}$ between sky lines.
Scaling our limiting sensitivity by $\sqrt{t}$, where $t$ is the integration time, we find it consistent with the quoted limits of other MOSFIRE $Y$-band observations reported by \citet{wirth15}. 

For each object in our mask, we computed the $X$-$\sigma$ limit in the rest-frame EW(Ly$\alpha$) as a function of redshift (i.e., observed wavelength) for our observations. This was done via 1000 Monte Carlo realizations of the photometry for each object, for which we performed SED fitting.
In each realization, the redshift was randomly drawn from the $p(z)$ distribution (thus the contaminant fraction, which is given by our $p(z)$, is accounted for in our results), and 
the corresponding continuum flux density redward of the Ly$\alpha$ was calculated from the best-fit SPS model.
The ratio of the limiting sensitivity, for which we take the median value at each wavelength as all the targets were observed in the same conditions in one MOSFIRE mask, to the continuum flux density in each realization gives the rest-frame $X$-$\sigma$ EW limit as a function of redshift (bottom left panel of Figure \ref{fig:simul}).

By assuming an intrinsic rest-frame EW distribution for Ly$\alpha$ before being processed by the neutral gas in the IGM, we can compute how many sources are expected to be detected above our $X$-$\sigma$ EW limit. 
For the intrinsic rest-frame EW distribution for Ly$\alpha$, $p$(EW$_{\rm intrinsic})$, 
we adopted a log-normal form given by \citet{schenker14}, which is based on the compilation of observations at $3<z<6$ when the universe is ionized.
Then, $p$(EW$_{\rm intrinsic})$ and our $X$-$\sigma$ EW limit inferred from our fake source simulation at the corresponding wavelength is compared, to estimate the probability that the line is detected. 
Here, we assumed that the $p$(EW$_{\rm intrinsic})$ does not evolve as a function of redshift from $3<z<6$ to $z=$ 7--8. Our analysis takes into account the effect of a sensitive wavelength dependancy due to sky lines and the incomplete spectral coverage of the redshift probability distribution ($p(z)$), and is properly weighted by $p(z)$.
The resulting probability distribution of the expected number of detections from our observations is shown in the right panel of Figure \ref{fig:simul}.
Depending on the detection threshold adopted, our results show a 1--2$\sigma$ deviation from the null hypothesis of no evolution.
For example, based on the Ly$\alpha$ EW distribution at lower redshift of $z \sim$ 3--6 (assuming no evolution with redshift), we expect to detect $1.7^{+0.6}_{-0.5}$ ($2.4^{+0.8}_{-0.8}$) objects with $> 5\sigma$ ($3\sigma$) significance, for which our observations weakly reject at the 1.3$\sigma$ (2$\sigma$) confidence level.
Our results are conservative in the sense that had we assumed a zero low-$z$ interloper fraction or used an extrapolation of the EW(Ly$\alpha$) distribution from lower redshifts to $z\sim$ 7--8, the inferred deviation from the expectation (and thus the implied decline in the Ly$\alpha$ fraction) would be higher.\footnote{
For reference, in a more traditional framework developed by \citet{stark10} of ``Ly$\alpha$ fraction'', 
z7\_GSD\_3811 is not regarded as a Ly$\alpha$-emitting galaxy, as the rest-frame EW is below the cutoff of 25 or 55 \AA.
Thus, the inferred Ly$\alpha$ fraction from our observation at $z \sim 7.5$ with EW $>$ 25 or 55 \AA\ is $X_{\rm Ly\alpha}
< 0.37$ for the UV-bright galaxies ($-21.75 < M_{\rm UV} < -20.25$) and $X_{\rm Ly\alpha} < 0.61$ for UV-faint galaxies ($-20.25 < M_{\rm UV} < -18.75$; $1\sigma$). 
}

\section{Discussion and Summary}\label{sec:discussion}

We have presented results from deep near-infrared $Y$-band spectroscopy targeting 12 galaxy candidates with $z_{\rm phot}=$ 7--8 in the GOODS-S field. 
Our long integration of $\sim$10 hr with Keck/MOSFIRE
enabled us to probe the Ly$\alpha$ emission down to a median 5$\sigma$ rest-frame EW(Ly$\alpha$) limit of 28 \AA\ (ranging [12--55] \AA; listed in Table \ref{tab:target}).  
Despite our deep spectroscopy, out of our 30 targets, we identified only one emission line at 6.5$\sigma$ significance.

We claim that the detection is real, given that $i$) it was detected independently on more than one night, $ii$) at the expected spatial location, and $iii$) with two negative peaks at the positions expected from our dithering pattern. 

This line is likely Ly$\alpha$ emission from a galaxy at $z=7.6637$, based on $i$) its asymmetric line profile characteristic of Ly$\alpha$ at high redshift, $ii$) the non-detection in the optical bands as well as an optical stack ({\it V$_{606}$}+{\it i$_{775}$}+{\it I$_{814}$}+{\it z$_{850}$} bands), and 
$iii$) the inferred redshift in good agreement with its photometric redshift.
While we cannot completely rule out the possibility that the detected line is an unresolved \OII\ doublet from a galaxy at $z=1.83$, we find that it is unlikely, as a serendipitious \OII\ emitter at $z \sim 1.9$ that falls in one of the slits, with the redshift difference of only $\Delta z \sim 0.1$ and with a similar S/N to that of z7\_GSD\_3811, shows clearly resolved double peaks both in our final stack and on individual nights. 
However, although rare, it is still feasible that the detected line is one of the two peaks of a broad \OII\ doublet indicative of an AGN or strong outflows.

The detected Ly$\alpha$ line has a modest rest-frame EW of 16 \AA\ and a line flux of $(5.5\pm0.9) \times 10^{-18}$ erg s$^{-1}$ cm$^{-2}$. This galaxy is bright in the UV ($M_{\rm UV}= -21.2$; $\sim 2L^*_{z=8}$), and is a typical for its UV brightness in terms of UV slope ($\beta=-2.2$) and stellar mass ($\log(M_*/M_{\odot})=9.3$).

Identifying its nature via follow-up observations would be challenging but not impossible.
Assuming this galaxy is an \OII\ emitter at $z=1.83$, other strong rest-frame optical emission lines (H$\beta$, \OIII, and H$\alpha$) all fall in between ground-based near-infrared bands, thus deep spaced-based grism may be the only possibility to detect those lines before the advent of the {\it James Webb Space Telescope}. 
If this galaxy is indeed at $z=7.6637$ (with a normal stellar population), other emission features (e.g., \CIII$\lambda\lambda$1907, 1909) 
would be too weak to be detected in currently available data sets (e.g., {\it HST} grism) based on the typical flux ratio, unless Ly$\alpha$ is attenuated more than a factor of 15 by the IGM. However, this is unlikely given the Ly$\alpha$ EW distribution found by \citet{stark11} and \citet{schenker14} for its UV luminosity in galaxies at $3<z<6$.
Additional integration in $Y$-band (for Ly$\alpha$) or deep $H$-band observations (for \CIII) can help verifying its identity.
Alternatively, the Atacama Large Millimeter Array (ALMA) provides an opportunity to detect the \CII\ line at 158 $\mu$m with less than an hour of integration, assuming that the empirical relation between SFR and \CII\ 158 $\mu$m luminosity found for normal star-forming galaxies at high redshift \citep{capak15} holds.
 
The rest of the targeted galaxies remain undetected, showing a $1.3\sigma$ ($2\sigma$) deviation from the expected number of detections (with $>$5$\sigma$ ($>$3$\sigma$) significance) when assuming no evolution in the Ly$\alpha$ EW distribution from lower redshifts of $3<z<6$ to $z=$ 7--8.
Our observations thus support the decline in the EW of Ly$\alpha$ at $z>6.5$ of earlier studies \citep[e.g.,][]{schenker14, tilvi14, pentericci14}, which may be due to the increase of neutral gas in the IGM. 
However, the evidence from our observations alone is not conclusive due to the large statistical uncertainties. The addition of our sample to the compilation of previous data would be only incremental, thus 
we defer a detailed analysis on the evolution of the IGM neutrality to future studies with a larger statistical sample. 

However, our results from very deep spectroscopy have implications for future observations.
Recently, several studies \citep{roberts-borsani15, zitrin15} have claimed a high Ly$\alpha$ visibility in bright galaxies at $z>7.5$ in the EGS field, which were selected based on red IRAC [3.6]$-$[4.5] colors indicative of strong \OIII\ emission. 
Combined with the recent discovery of Lyman continuum leakers among strong \OIII\ emitters at low redshifts \citep{izotov16,vanzella16} and the lack of significant Ly$\alpha$ detections in the GOODS-S field at comparable redshifts (before this study), this may signal the inhomogeneity of the reionization process on large scales.
Indeed, while LAEs at lower redshifts of $3<z<6$ show that faint LAEs on average have a larger Ly$\alpha$ EW than bright ones \citep{stark11}, most spectrosopic campaigns at higher redshift targeting $z>7$ galaxy candidates have only succeeded in detecting Ly$\alpha$ emission in bright galaxies \citep{finkelstein13,oesch15,roberts-borsani15,zitrin15}. 
Our results are in line with these studies, yielding one Ly$\alpha$ detection from a bright ($L \sim 2L^*$) galaxy.
However, it is noteworthy that our sole detection in z7\_GSD\_3811 is among those with the lowest EW limit (cyan circles in the bottom left panel of Figure \ref{fig:simul}). 
This indicates that current spectroscopic campaigns at $z>7$ are only reaching a sufficient depth for the brightest galaxies, leaving the possibility of detecting several galaxies in Ly$\alpha$ emission with modest Ly$\alpha$ EW in fainter galaxies open with deeper spectroscopy.
Extremely deep spectroscopy (either by performing long integrations on blank fields or by utilizing magnification due to gravitational lensing)
to better quantify the Ly$\alpha$ EW distribution, along with quantifying large-scale spatial fluctuation in the reionization process from spatial clustering of Ly$\alpha$ emission from wide area surveys, will remain as a valuable probe of reionization in the near future.

\acknowledgements
\noindent
We would like to thank the anonymous referee for valuable suggestions which improved this paper, and N. Scoville, L. Murchikova, S. Manohar, and B. Darvish for their work during observing runs.
M.S. acknowledges support from the NSF AAG award AST-1518183.  M.S. and S.L.F. acknowledge support from the NASA Astrophysics and Data Analysis Program award \#NNX15AM02G issued by JPL/Caltech, as well as a NASA Keck PI Data Award, administered by the NASA Exoplanet Science Institute. 
This research is based on observations made with the Keck Telescope. The Observatory was made possible by the financial support of the W. M. Keck Foundation. We recognize and acknowledge the cultural role and reverence that the summit of Mauna Kea has within the indigenous Hawaiian community.

~

{\it Facilities:}  \facility{Keck:I (MOSFIRE)}, \facility{{\it HST} (ACS, WFC3)}, \facility{VLT (HAWK-I)}


\end{document}